DRAFT

# QUANTIFICATION OF ALVEOLAR RECRUITMENT FOR THE OPTIMIZATION OF MECHANICAL VENTILATION USING QUASI-STATIC PRESSURE VOLUME CURVE

Mohsen Nabian, Uichiro Narusawa

Department of Mechanical and Industrial Engineering
Northeastern University, Boston, MA

**ABSTRACT**

Quasi-static, pulmonary pressure-volume (P-V) curves over an inflation-deflation cycle are analyzed using a respiratory system model (RSM), which had been developed for quantitative characterization of the mechanical behavior of the total respiratory system. Optimum mechanical ventilation setting of Positive End Expiratory Pressure (PEEP) for total alveolar recruitment is quantified based on the existing P-V curves of healthy and injured animal models. Our analytical predictions may contribute to the optimization of mechanical ventilation settings for the Acute Respiratory Distress Syndrome (ARDS) patients.

**INTRODUCTION**

Quasi-static pulmonary pressure-volume (P-V) curves are used for clinical guidance to obtain quantitative information on the mechanical behavior of the respiratory system as well as the conditions of gas exchange [1-12]. There have been extensive analytical and numerical research work with the goal of providing decent respiratory models to enhance our understanding of healthy and injured lungs and further improving the concurrent mechanical ventilation procedures for people with lung injury [3-5,13-16]. Mechanical ventilation is an effective and often the only solution in the care of patients with acute respiratory distress syndrome (ARDS) [6,17-24].
Mechanical ventilation parameters such as positive end-expiratory pressure (PEEP) and Tidal Volume are the inputs parameters of mechanical ventilation. There are extensive experimental and theoretical works with the aim of finding the optimum values for these parameters. Some protective ventilation strategies have been proposed for patients in intensive care units. For example, positive end-expiratory pressure (PEEP) has been associated with alveolar recruitment for effective gas exchange as well as lung over-distention which results in ventilator-induced lung injury [25-36]. However, it is a difficult task to propose optimum conditions of ventilation in terms of both the efficacy and the safety of the entire ARDS patient populations given large variability in the mechanical properties of the respiratory systems in the ARDS patients. Therefore, integration of personal data with proper modeling of respiratory system may contribute to the patient-specific optimization of mechanical ventilation.

In this paper, using a respiratory system model (RSM) [4,5,6], we developed equations for estimating the volume recruitment of respiratory system during mechanical ventilation as functions of respiratory system model parameters as well as the ventilation Positive End Expiratory Pressure (PEEP). The respiratory system model parameters can be obtained by fitting the RSM to the patient's quasi-static pressure-volume curves. Therefore, these patient-specific parameters are related to the patient's overall mechanical properties of the respiratory system.

In this paper, we applied our analysis to the pressure-volume data of healthy and canine models of ARDS. We quantitatively demonstrated that although the zero end expiratory pressure (ZEEP) lead to maximizing the volume recruitment of healthy respiratory system, large positive end expiratory pressure (PEEP) is required for ARDS models to achieve maximum volume recruitment. Many previously published experimental works reported similar results as our analytical model [6,22,25,26,27,29,32,34]. Our findings from dog ARDS models may be directly applied for the optimization of mechanical ventilation of human ARDS patients.

**Nomenclature**

| | |
|---|---|
| $A_s$ | piston surface area on which pressure is acting (Fig.2) |
| $f(F)$ | (normalized) inflation distribution function in Eq.(2) |
| $I_i$ | See Eq.(3) for definitions. |
| $k$ | spring constant [N/m] (Fig.2) |
| $N$ | total number of alveolar elements of total respiratory system. |
| $N_j$ | number of alveolar elements of total respiratory |
| $V$ | Volume |
| $V_{U(L)}$ | Upper (lower) volume asymptote (fig.1) |
| $\bar{V}$ | $=[V-(V_U-\frac{\Delta V}{2})]/(\Delta V/2)$ |
| $\Delta V$ | $=V_U-V_L \ (Fig.1)$ |
| $\hat{V}_0$ | 'pop-open' volume of an element in inflation $(=A_s\hat{y}_0)$ |
| $\hat{y}_0$ | Stroke limit of the piston due to alveolar opening. |
| $\hat{y}_T$ | Stroke limit of the piston due to tissue distension. |
| $\hat{y}_{T0}$ | $=\hat{y}_T/\hat{y}_0$ |
| $\Lambda$ | Non-dimensional inflation parameter |
| $\sigma_D$ | Standard deviation of the normal distribution Eq.(2) |
| ARDS | Acute respiratory distress syndrome |
| DG | data group |
| FRC | Functional residual capacity |
| LIP | Lower inflection point |
| PEEP | Positive expiratory pressure |
| RSM | Respiratory system model |
| TLAIFRC | Tidal loop after inflation from FRC |
| TLADTLC | Tidal loop after deflation from TLC |
| TLC | Total lung capacity(=$V_U$) |
| UIP | Upper inflection point |
| VLC | Vital lung capacity(=$\Delta V$) |
| ZEEP | Zero end expiratory pressure |

## SUMMARY OF THE PREVIOUS WORK

A P-V model equation with its parameters determined from statistical processing of P-V data points quantifies characteristics of P-V curves as well as their changes observed in clinical settings. Fig.1 (a) shows a typical P-V curve during inflation and deflation. Inflation P-V curves may be expressed in a P-V model equation as

$$V = V_U - \frac{\Delta V}{2} + \frac{\Delta V}{2} \cdot \text{erf}\left(\frac{\sqrt{\pi}}{4}\Lambda \cdot \left(\frac{P}{P_0} - 1\right)\right) \quad (1)$$

where four parameters are $V_U$ = the upper asymptote, total lung capacity (TLC) (or $V_L$ = the lower asymptote, functional residual capacity (FRC)), $\Delta V$ (=$V_U - V_L$) = vital lung capacity (VLC), $P_0$ = pressure at the midpoint (inflection point) of the P-V curve, and $\Lambda$ = non- dimensional parameter related to the slope of the P-V curve at P = $P_0$. The error function, $erf(x)$, is closely related to the normal (bell-shaped) distribution through an integral, defined as $erf(x) = (2/\sqrt{\pi})\int_0^x e^{-t^2}dt$, $erf(\infty) = 1$ and $erf(-x) = erf(x)$.

The deflation P-V curves are also represented by Eq.(1) with the four inflation parameters replaced by the corresponding deflation parameters of $\Lambda^d, V_U^d, \Delta V_U^d, P_0^d$ (See Fig.1 (a) a typical P-V curve, indicating the model parameters and Fig.1(b) for various inflation-deflation loops for 2 dogs before and injury.)

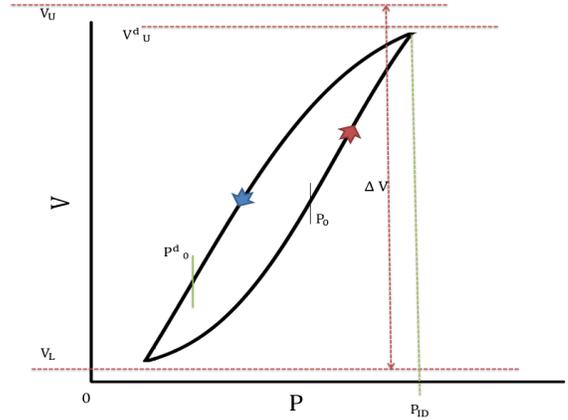

**Fig.1(a)** a typical quasi-static P-V curve of a dog pulmonary system.

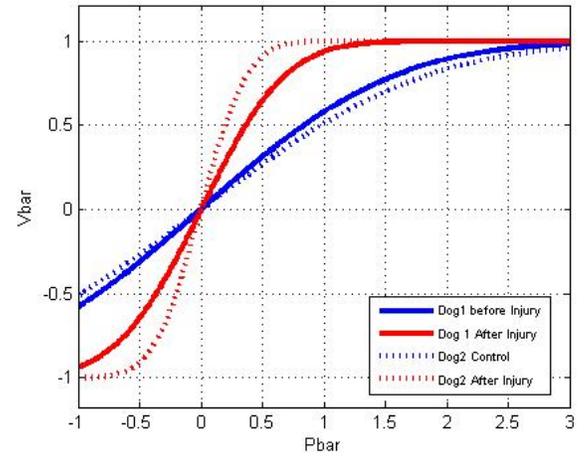

**Fig.1(b)** non-dimensional pressure-volume curve of Dog1 and Dog2 models before and after injury

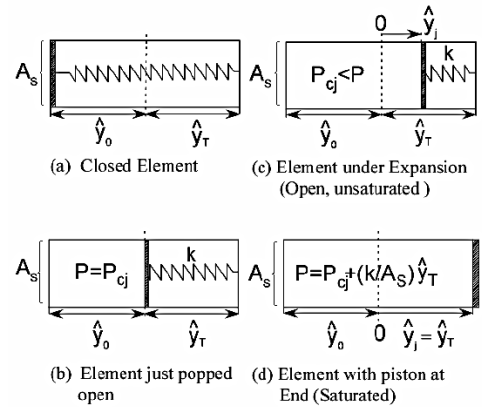

**Fig 2** Alveolar element for inflation (left)

Presented below is an outline of RSM for the inflation process [4]. In the model, a total respiratory system is represented by a large population of basic alveolar elements (N

= total number of elements) distributed over its (alveolar) opening pressure $P_{cj}$. Referring to Fig.2 (a), an alveolar element, j, consists of a cylindrical chamber with a piston-spring system ($A_s$ = piston surface area, $k$ = spring constant [N/m], $\hat{y}_0$= piston displacement due to pop-open mechanism, $\hat{y}_T$= piston displacement due to elastic wall distension). The element is closed when the piston is located at the left end of the cylinder. When the pressure acting on the left hand side of the piston reaches $P_{cj}$, the piston moves to a new position ('pop-open' mechanism = alveolar recruitment) with $\widehat{V_0} = (A_s\widehat{y_0})$ ) indicating an elemental volume increase due to the piston displacement of $\widehat{y_0}$. (see figure 2 (a), (b)).

Any further change in pressure results in a volume increase as the piston moves to the right ($0 < \hat{y}_j < \widehat{y_T}$) (Simulating the elastic distension of the alveolar wall tissues) until it reaches the end of the cylinder ( $y_j = \widehat{y_T}$) as a state of fully-distended element. An application of the Boltzmann statistical model yields a normal distribution as the most probable distribution of alveolar elements over their opening pressure, $P_{cj}$, $dN_j/N$ ( = a number fraction of elements, for which the magnitude of $P_{cj}$ ranges between $P_{cj}$ and $P_{cj} + dP_{cj}$),

$$\frac{dN_j}{N \cdot dP_{cj}} = f(P) \qquad (2)$$

$$f(P) = \frac{\Lambda}{4P_0} \cdot \exp\left(-\left(\frac{\sqrt{\pi}}{4}\Lambda\right)^2 \left[\frac{P_{cj} - (P - P_0)}{P_0}\right]^2\right)$$

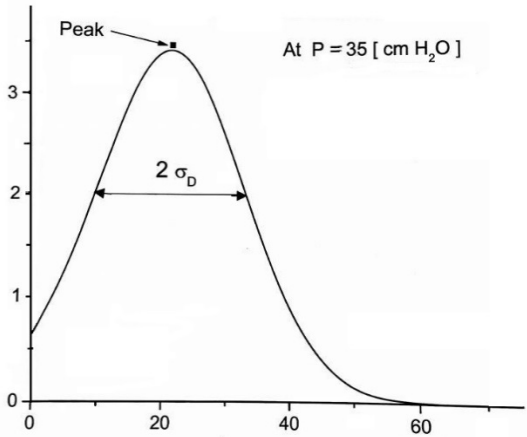

**Fig.3** Alveolar Opening Pressure, $P_{cj}$ [cm H$_2$O]

Fig.3 sketches a normal distribution of alveolar opening pressure $P_{cj}$ (Eq.(2)) at the inflation pressure of $P = 35$ [cmH2O]. The area under the distribution from $-Infinity$ to 35 [cmH2O] represents the number of alveolar elements that are open at $P = 35$ [cmH2O]. The distribution peak value is $\Lambda/4P_0$, located at $P_{cj} = 35$ [cmH2O] - $P_0$ with the standard deviation, $\sigma_D$ (= $(8/\pi)^2 P_0/\Lambda$) $\approx 1.596 \cdot P_0/\Lambda$. By summing (integrating) the volume of each element over all open elements RSM yields two P-V equations, one for the low-P region and the other for the high-P region [4]. The equation for the high-P region is the error function P-V equation (= Eq.(1)). However, Eq.(1) with the four adjusting parameters has been shown to be an excellent approximation for both inflation- and deflation P-V curves over the entire P-region [5,6]. Also, RSM gives the following equation for $V_U^d$ in terms of parameters of the inflation preceding the deflation;

$$V_U^d = V_U - \Delta V + \frac{2\Delta V}{\pi\Lambda(1+\hat{y}_{TO})} \cdot I_5 + \frac{\Delta V}{2} \cdot [1 + I_3(\bar{p}_{ID})] + \frac{\Delta V(1-\hat{y}_{TO})}{2(1+\hat{y}_{TO})}(I_4 - I_1) \qquad (3)$$

Where $I_5 = \exp(-C^2(\hat{y}_{TO} - 1)^2) - \exp(-C^2)$ with $= \sqrt{\pi}\frac{\Lambda}{4}$, $I_3(p) = \mathrm{erf}(C(\frac{p}{p_0} - 1))$, $I_4 = \mathrm{erf}(C(1 - \hat{y}_{TO}))$, $I_1 = \mathrm{erf}(C)$.

Eq.(3) is used to obtain the magnitude of $\widehat{y_{TO}}$ from the P-V curve. Eq.(1), expressed in terms of non-dimensional volume ($\bar{V}$) and non-dimensional pressure ($\bar{P}$) [41], is,

$$\bar{V} = \mathrm{erf}\left(\frac{\sqrt{\pi}}{4}\Lambda \cdot \bar{P}\right) \qquad (4)$$

Where

$$\bar{V} = \frac{V - \left(V_U - \frac{\Delta V}{2}\right)}{\frac{\Delta V}{2}}, \qquad \bar{P} = \frac{P}{P_0} - 1$$

Also, Eq.(2) may be transformed into the following normalized form;

$$\frac{dN_j}{N \cdot d\widehat{P}_{cj}} = f(\bar{P})$$

in which

$$f(\bar{P}) = \frac{\Lambda}{4} \cdot \exp\left(-\left(\frac{\sqrt{\pi}}{4}\Lambda\right)^2 \left[\widehat{P}_{cj} - \bar{P}\right]^2\right) \qquad (5)$$

where $\widehat{P}_{cj} = P_{cj}/P_0$ with the standard deviation, $\sigma = (8/\pi)^2/\Lambda$ ($\approx 1.596/\Lambda$). It should be noted that the normalized P-V equation as well as the normalized distribution depend on a single non-dimensional parameter, $\Lambda$. Table 1 summarizes P-V parameters and their description in Fig.1(a) and in Eqs.(1)(2) as well as the corresponding non-dimensional parameters and $\bar{P} - \bar{V}$ locations in Eqs.(4)(5).

By multiplying the normal distribution of the alveolar elements (= Eq.(2)) by $\widehat{V}_0 = A_s\hat{y}_0$, the elemental volume recruited by the pop-open mechanism [4]) and noting $N\widehat{V_0}(1 + \widehat{y_{TO}})$ with $\hat{y}_{TO} = \hat{y}_T/\hat{y}_0$, the resulting distribution is a recruitment volume distribution expressed as,

$$\widehat{V}_0 \cdot \left(\frac{dN_j}{dP_{cj}}\right) = \left(\frac{\Delta V}{1 + \hat{y}_{TO}}\right) \cdot f(P) \qquad (6)$$

**Table 1**. P-V locations and the corresponding $\bar{P} - \bar{V}$

| Equations (1),(2) | Description | Equations (3),(4) |
|---|---|---|
| $V_U$ | Upper asymptote | $\bar{V} = 1$ |
| $V_L$ | Lower asymptote | $\bar{V} = -1$ |
| $(V_U+V_L)/2$ | V at inflation point | $\bar{V} = 0$ |
| $\Delta V$ |  | $\Delta \bar{V} = 2$ |
| $P = 0$ |  | $\bar{P} = -1$ |
| $P_0$ | P at inflation point | $\overline{P_0} = 0$ |
| $\frac{\Lambda}{4P_0} \cdot \Delta V$ | Compliance at inflection point | $\frac{\Lambda}{2}$ |
| $\frac{\Lambda}{4P_0}$ | Peak value of the normal distribution | $\frac{\Lambda}{4}$ |
| $\sigma_D = \sqrt{8/\pi}P_0/\Lambda$ | Standard deviation of the normal distribution | $\sigma = \sqrt{8/\pi}/\Lambda$ |

Sources of data sets analyzed and quoted herein are;
1. Data Group 1 (DG1) [59]: Canine model of acute lung injury before and after oleic acid-induced injury (DG1 B.I. and DG1 A.I.).
2. Data Group 2 (DG2) [17]: Canine model of acute lung injury by saline lavage; Digitized P − V curves (control) (DG2 Ctrl.) vs after lung injury (DG2 A.I.)).

Table 2, 3 lists the P-V curve parameters of DG1 and DG2, found by fitting Eq.(1) to data points by the method of least squares.

**Table 2(a).** Inflation P-V curve parameters.

|  |  | $\Lambda$ | $P_0$ | $\Delta$ | $V_U$ | $\widehat{y_{T0}}$ |
|---|---|---|---|---|---|---|
| Dog 1 | Before Injury | 1.25 | 12.57 | 1.12 | 0.88 | 0.22 |
|  | After Injury | 2.96 | 20.79 | 0.86 | 0.83 | 0.84 |
| Dog 2 | Control | 1.10 | 8.36 | 1.50 | 1.16 | 0.10 |
|  | After Injury | 5.40 | 22.29 | 1.12 | 1.10 | 0.76 |

**Table 2(b).** Deflation P-V curve parameters.

|  |  | $\Lambda^d$ | $P_0^d$ | $\Delta V^d$ | $V_U^d$ |
|---|---|---|---|---|---|
| Dog 1 | Before Injury | 0.28 | 2.0 | 1.43 | 1.03 |
|  | After Injury | 0.40 | 3.8 | 1.11 | 0.78 |
| Dog 2 | Control | 0.22 | 1.3 | 2.02 | 1.12 |
|  | After Injury | 1.45 | 9.8 | 1.19 | 1.05 |

## TIDAL LOOP ANALYSIS : TLAIFRC AND TLADTLC

The artificial ventilation strategy must be optimized carefully by taking into account the degree of severity in lung injury. Analyses are made on cyclic inflation-deflation loops (to be referred to as tidal loops, simulating artificial ventilation) between a high pressure (Peak) and a low pressure (PEEP (ZEEP) = positive (zero) end-expiratory pressure) along a specified envelope P-V curve of inflation and deflation. There are two modes of small volume tidal loop for ventilation in relation to the main (inflation and deflation) P-V curves [Downie, Nam and Simon [4]]. They are Tidal Loops performed After Inflation from FRC (TLAIFRC) and After Deflation from TLC (TLADTLC) with FRC = functional residual capacity and TLC = total lung capacity. In this study we followed by a tidal inflation limb and then a tidal deflation limb over the tidal pressure of $\Delta P_{tidal} = $ Peak - PEEP;
Use V ($P_{tp} = P_{ID} \approx 30$cm $H_2O$) as TLC. The two modes are respectively sketched in Fig.4 and 5. In TLAIFRC Fig.4, after inflation from P = 0 to a certain (tidal) peak pressure, the tidal loop begins with a main tidal deflation limb, that is, in Fig.4:
[O → A → B] = main inflation, [B → C] = main tidal deflation,
[C → D] = tidal inflation 1, [D → E] = tidal deflation 1,
[E → F] = tidal inflation 2, [F → G] = tidal deflation 2.

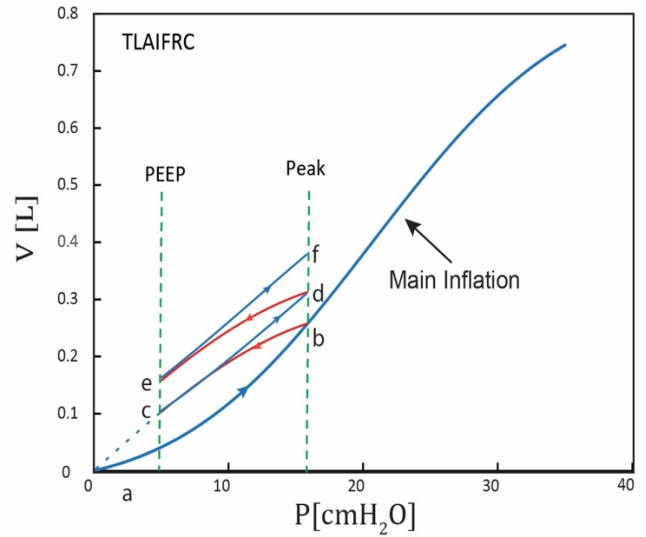

**Fig4.** Pressure-Volume curve for Tidal loop after inflation from FRC (TLAIFRC)

In TLADTLC (Fig.5), on the other hand, after deflation from a certain high pressure, PID (regarded as the pressure at TLC) preceded by a main inflation process, the first tidal loop is initiated as a tidal inflation, followed by a tidal deflation; in Fig.5:

[O → a → b → ID] = main inflation
[ID → c → d] = main deflation
[d → e] = tidal inflation 1, [e → f ] = tidal deflation 1,
[f → g] = tidal inflation 2, [g → h] = tidal deflation 2.

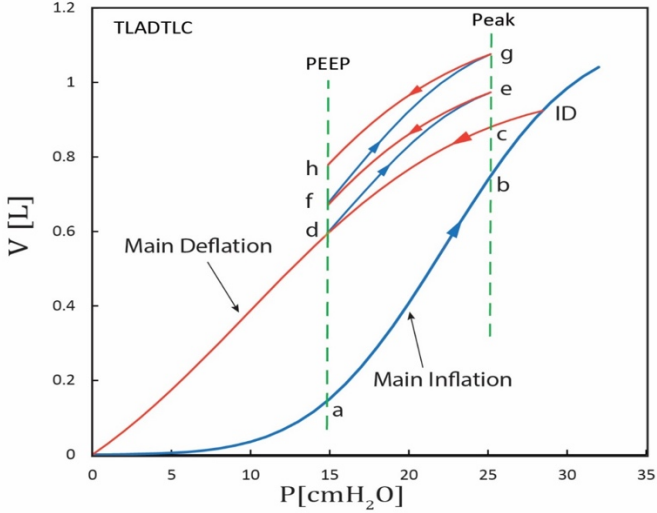

**Fig5.** Pressure-Volume curve for Tidal loop after deflation from TLC (TLADTLC)

Table 3(a) describes states of alveolar elements participating in TLAIFRC. They are classified into four categories, shown respectively in the four columns. The first column is on elements with their alveolar opening pressure between P = 0 and PEEP (0 → PEEP) in inflation and also with their closing pressure between P = 0 and PEEP (0 ⇐ PEEP) in deflation. These elements are recruited during the process from 0 → A along the main inflation in Fig.4. They remain open (recruited) throughout the tidal loops between Peak and PEEP because they may be derecruited only in the deflation process from PEEP to P = 0; thus not contributing to recruitment/derecruitment in the tidal loops to follow. On the other hand, elements in the second column (to be recruited in inflation from P = 0 to PEEP, and derecruited in deflation from Peak to PEEP) may be recruited along the main inflation (0 → A) and derecruited along the main tidal deflation (B → C), only to remain closed in the ensuing tidal loops as these loops do not cover the pressure range for recruitment between P = 0 and PEEP. It is the type 4 elements satisfying the conditions of the last column that fully participate in tidal loops as both their opening and closing pressures are within the tidal pressure range of PEEP ↔ Peak. Table 3(b) shows states of elements in TLADTLC sketched in Fig.5. An argument similar to the one above for TLAIFRC indicates that the type 4 elements in Table 3(b) respond to the tidal pressure change as they move between open and closed states for recruitment-derecruitment cycles.

In order to find magnitudes of these four parameters, there exist three equations that relate a deflation limb with the preceding inflation. One is the volume equality at P = Peak. The second is $V^d_{U,TDL}$ in terms of the inflation parameters (Eq.(3)). The third relation specifies the volume of tidal deflation at P = 0 ($V^d_{TDL}$ (P = 0) = specified). Also, as a first order approximation, the magnitude of $\Lambda$ is assumed to be independent of Peak pressure (i.e. $\Lambda^d_{TDL} = \Lambda_d$). This approximation implies that, for a specified respiratory system, the normalized distribution for the main tidal deflation is similar in shape for all main tidal deflation curves (i.e. $\Lambda^d_{TDL}$ does not depend on $P_{ID}$). Using these relations, the parameters of the main tidal deflation curve are determined for each case of TLAIFRC with a different Peak. Shown in Fig.6 are the main inflation P-V curve as well as main tidal deflation curves (α, β, γ, δ), each starting at a different Peak value for the data set of DG2 After Injury. The PEEP locations are marked for the case of $\Delta P_{tidal}$ = 10[cmH2O]. For example, a part of the α curve from P = 15[cmH2O] (Peak) to P = 5[cmH2O] (PEEP) is the main tidal deflation curve after inflation from P = 0 to 15 [cmH2O] for TLAIFRC.

According to RSM, the quasi-static P-V curve of the inflation

**Table 3(a).** States of elements in TLAIFRC of figure 4

| Type | 1 | 2 | 3 | 4 |
|---|---|---|---|---|
| P-range for recruitment | PEEP ⇑ 0 | PEEP ⇑ 0 | Peak ⇑ PEEP | Peak ⇑ PEEP |
| P-range for derecruitment | 0 ⇑ PEEP | PEEP ⇑ Peak | 0 ⇑ PEEP | PEEP ⇑ Peak |
| After Main Inf., 0→B | Open | Open | Open | Open |
| After Main Tidal Def., B→C | Open | Close | Open | Close |
| After Tidal Inf.1, C→D | Open | Close | Open | Open |
| After Tidal Def.1, D→E | Open | Close | Open | Close |
| After Tidal Inf.2, E→F | Open | Close | Open | Open |

(deflation) process is related to a distribution of the alveolar elements over their opening (closing) pressure (Eq.(2)). Therefore, referring again to the α curve in Fig.8, alveolar elements with the range of their opening pressure between 0[cmH2O] and 15[cmH2O] are recruited along the main inflation curve, followed by the main tidal deflation along the curve α, in which the elements with their alveolar closing (collapsing) pressure between 15[cmH2O] and 5[cmH2O] are derecruited. A theoretical prediction of recruited- or derecruited- volume in a tidal pressure loop may be made from the distribution of elements over the alveolar opening/closing *pressure*. The basic equation is the one for a number of alveolar elements as a fraction of the total number of elements (number fraction (NF)) that are recruited during an inflation process from $P = P_i$ to $P = P_e$ ( $>P_i$) ( $\equiv NF_{rec}$.( $P_i \rightarrow P_e$) ).

It is obtained by integrating the distribution function over the range of the critical pop-open pressure of $P_i \leq P_{cj} \leq P_e$: A similar equation may be derived for the deflation process; i.e.

**Table 3(b).** States of elements in TLADTLC of figure 5

| Type | 1 | 2 | 3 | 4 | 5 | 6 | 7 |
|---|---|---|---|---|---|---|---|
| P-range for Recruitment | PEEP ⇑ 0 | PEEP ⇑ 0 | Peak ⇑ PEEP | Peak ⇑ PEEP | Peak ⇑ PEEP | ID ⇑ Peak | ID ⇑ Peak |
| P-range for Derecruitment | 0 ⇑ PEEP | PEEP ⇑ ID | 0 ⇑ PEEP | PEEP ⇑ Peak | Peak ⇑ ID | 0 ⇑ PEEP | PEEP ⇑ ID |
| After Main Inf., 0→B | Open | Open | Open | Open | Open | Open | Open |
| After Main Tidal Def., B→C | Open | Close | Open | Close | Close | Open | Close |
| After Tidal Inf.1, C→D | Open | Close | Open | Open | Open | Open | Close |
| After Tidal Def.1, D→E | Open | Close | Open | Close | Open | Open | Close |
| After Tidal Inf.2, E→F | Open | Close | Open | Open | Open | Open | Close |

the number fraction of elements derecruited during a deflation process from $P = P_{id}$ to $P = P_e^d (<P_i^d)$ : The alveolar opening pressure $(P_{cj})$ of the inflation and the alveolar closing pressure (of the deflation are independent of each other as random variables of the inflation- and the deflation-distribution [Reza-Inf/def]. Hence, the number fraction (or probability) of alveolar elements that are recruited during the inflation from $P_i$ to $P_e$, and subsequently derecruited during the deflation from $P_i^d$ to $P_e^d$ may be expressed in a joint distribution represented by a product of the inflation and the deflation distribution (probability function).

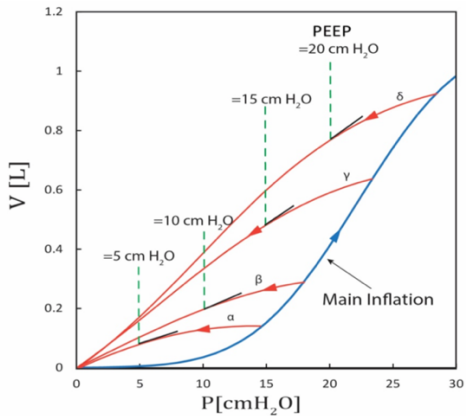

**Fig6**, Relation between main inflation and tidal inflation

For the case of TLAIFRC sketched in Fig.4(a), we define
$N (C \to D)$ = number of elements to be recruited in tidal inflation process of $C \to D$,
$N(O \leftrightarrow B)$ = number of elements participating in the main inflation - main tidal deflation loop of $O \leftrightarrow B$, then, $N (C \to D)$ as a joint distribution of inflation and deflation becomes,

$$N(C \to D) \equiv \left\{ \begin{array}{c} \text{No. of recruitable elements} \\ \text{in tidal inflation process f} \to \text{g} \end{array} \right\}$$

$$= N(O \leftrightarrow B) \cdot \frac{NF_{rec.}(A \to B)}{NF_{rec.}(O \to B)} \cdot \frac{NF_{derec.}(B \to C)}{NF_{derec.}(B \to O)} \quad (7)$$

The second and the third factor on the right-hand side of Eq. (7) are respectively a number of elements recruited along the main inflation $A \to B$ as a fraction of elements participating in the P-V loop of $O \leftrightarrow B$, and a number of elements derecruited along the main tidal deflation $B \to C$ as a fraction of elements participating in the P-V loop of $O \leftrightarrow B$. Therefore:

$$\Delta V_{Rec,TLAIFRC} \equiv \left\{ \begin{array}{c} \text{Volume to be recruited in tidal inflation} \\ \text{process from PEEP to Peak} \end{array} \right\}$$

$$= \hat{V}_0 \cdot N(C \to D) \quad (8)$$

For the case of TLADTLC sketched in Fig.5, elements participating in the tidal inflation from State f (= PEEP) to State g (= Peak) are those that are recruited along the main inflation in a (PEEP) → b (Peak) and subsequently derecruited along the main deflation between c (Peak) and d,

$$N(f \to g) \equiv \left\{ \begin{array}{c} \text{No. of recruitable elements} \\ \text{in tidal inflation process f} \to \text{g} \end{array} \right\} =$$

$$= N(O \leftrightarrow ID) \cdot \frac{NF_{rec.}(a \to b)}{NF_{rec.}(O \to ID)} \cdot \frac{NF_{derec.}(c \to d)}{NF_{derec.}(ID \to O)} \quad (9)$$

The second and the third factor on the right-hand side of Eq.(9) are respectively a probability of elements recruited along the main inflation a $\to$ b, and a probability of elements derecruited along the main deflation c $\to$ d, both as a fraction of elements participating in the main P-V loop of $O \leftrightarrow ID$. Therefore,

$$\Delta V_{Rec,TLADTLC} \equiv \left\{ \begin{array}{c} \text{Volume to be recruited in tidal inflation} \\ \text{process of PEEP} \to \text{Peak} \end{array} \right\} =$$

$$= \hat{V}_0 \cdot N(f \to g) \quad (10)$$

In the following figure (Fig.7), the volume recruitment of tidal inflation in both TLADTLC and TLAFRC methods over a wide range of Peak pressure is demonstrated. For this analysis, ΔP=10 cmH$_2$O is constant while peak pressure varies between 10 to 35 cmH$_2$O. The TLADTLC and TLAIFRC for both healthy and injured samples are presented as continuous lines and dot lines respectively.

According to figure.7, the volume recruitment in healthy lungs is constantly decreasing by the increment of Peak pressure in tidal inflation, which implies that choosing ZEEP (Zero end expiratory pressure) as the tidal end expiratory pressure would maximize the lung volume recruitment.

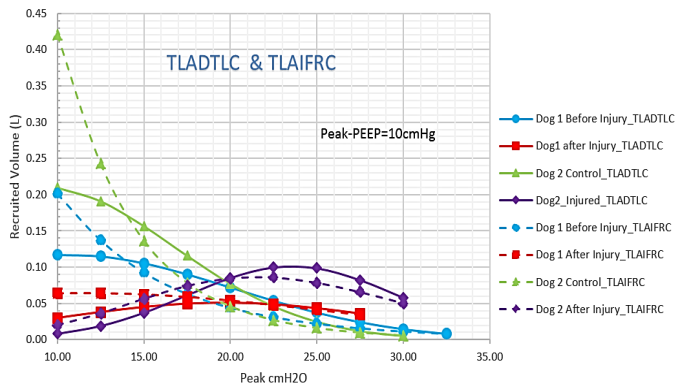

**Fig7**. Volume recruitment of tidal inflation in *TLADTLC* and *TLAFRC* over a wide range of Peak pressure value with $\Delta P=10$

However, interestingly demonstrated in Fig.7, for the injured lungs, a maximum volume recruitment occurs in PEEP other than 0 which supports many experimental research work suggesting positive end expiratory pressure for the optimum mechanical ventilation [6,22,25,26,27,29,32,34]. This optimum PEEP is quantitatively proposed in Fig.7 for the two injured dog respiratory systems.

It is not possible to prioritize TLADTLC ventilation method over TLAFRC or vice-versa based on mere alveolar recruitment analysis (Fig 7), however, this analysis, proposes that the optimum pressure for PEEP in TLADTLC method is slightly higher than the optimum PEEP in TLAFRC.

## SUMMARY AND CONCLUSION

In this study, an analytical analysis of lung recruitment is performed based on previously introduced RSM method which quantifies the alveolar opening pressure with statistical distribution function. This study further demonstrates that among all alveoli, the elements with their alveolar opening pressure lying between the tidal pressures will respond to the tidal pressure change as they move between open and closed states for recruitment-derecruitment cycles. This activated volume known as recruitment volume is quantified over a large tidal pressure for two healthy and injured dog models. This analysis quantitatively demonstrates that a positive end expiratory pressure would maximize alveolar recruitment in injured dogs. This is in contrast for the healthy dogs that zero end expiratory pressure will maximize volume recruitment in mechanical ventilation. Many previously published experimental works reported similar results as our analytical model. Our findings from dog ARDS models may be applied for the optimization of mechanical ventilation of human ARDS patients.